\documentclass[10pt,conference]{IEEEtran}

\usepackage{amsmath}
\usepackage{subfig}
\usepackage{comment}
\usepackage[noend]{algpseudocode}
\usepackage[T1]{fontenc}
\usepackage{lmodern}
\usepackage{multirow}
\usepackage{amsmath}
\usepackage{mathtools}
\usepackage{xcolor}
\usepackage[ruled, vlined, linesnumbered]{algorithm2e}
\usepackage{amssymb}
\usepackage{bbding}
\usepackage{pifont}
\usepackage{tikz}

\usepackage[compact]{titlesec}         
\titlespacing{\section}{2pt}{2pt}{2pt} 
\AtBeginDocument{
  \setlength\abovedisplayskip{2pt}
  \setlength\belowdisplayskip{2pt}}

\usepackage[rawfloats=true]{floatrow} 
\restylefloat{figure}     
\ifCLASSOPTIONcompsoc
  \usepackage[nocompress]{cite}
\else
  \usepackage{cite}
\fi

\usepackage{graphicx}
\ifCLASSINFOpdf
  
\else
 
\fi

\usepackage[rawfloats=true]{floatrow} 
\begin{document}
%

\title{Synergistic Integration of Blockchain and Software-Defined Networking in the Internet of Energy Systems}
\author{Vahideh Hayyolalam\IEEEauthorrefmark{1}, Abdulrezzak Zekiye\IEEEauthorrefmark{1}, Hamza Abuzahra\IEEEauthorrefmark{1}, 
Öznur Özkasap\IEEEauthorrefmark{1}\\
Murat Karakus\IEEEauthorrefmark{2}, Evrim Guler\IEEEauthorrefmark{3}, Suleyman Uludag\IEEEauthorrefmark{4} \\ 
\IEEEauthorrefmark{1}Department of Computer Engineering, Koç University, İstanbul, Türkiye \\
\IEEEauthorrefmark{2}Department of Software Engineering, Ankara University, Ankara, Türkiye \\
\IEEEauthorrefmark{3}Department of Computer Engineering, Bartin University, Bartin, Türkiye \\
\IEEEauthorrefmark{4}Department of Computer Science, University of Michigan - Flint, MI, USA \\
Emails: \IEEEauthorrefmark{1}\{vhayyolalam20, azakieh22, habuzahra24, oozkasap\}@ku.edu.tr, \\
\IEEEauthorrefmark{2}mrtkarakus@ankara.edu.tr, 
\IEEEauthorrefmark{3}evrimguler@bartin.edu.tr, 
\IEEEauthorrefmark{4}uludag@umich.edu}
%

%
%





\maketitle

\begin{abstract}
    Peer-to-peer (P2P) energy trading, Smart Grids (SG), and electric vehicle energy management are integral components of the Internet of Energy (IoE) field. The integration of Software-Defined Networks (SDNs) and Blockchain (BC) technologies into the IoE domain offers potential benefits that have only been studied in the literature in a few works. In this paper, we investigate the state-of-art solutions that leverage both SDNs and blockchain within the realm of the IoE. We categorize these solutions based on the method of integrating SDN and BC into two categories. The first category is the blockchain for SDN, where blockchain enhances the SDN directly. The second category is blockchain and SDN, where both technologies are used to enhance the proposed solutions. We identify three distinct blockchain applications based on their usage: decentralizing the SDN control plane, serving as a decentralized platform, and improving security measures. Similarly, we observe that SDN serves as a performance enhancer, a substitute for traditional networking, and solely as a control and management framework. It is posited that integrating SDNs and blockchain into IoE leads to performance enhancements, improves security, enables decentralized operations, and eliminates single points of failure in the SDN control plane. Additionally, some unaddressed issues, such as energy efficiency, smart contract management, and scalability, are discussed as potential future directions.
\end{abstract}

\begin{IEEEkeywords}
Blockchain, Internet of Energy, SDN, Smart Grids, Electric Vehicles. \vspace{-.4cm}

\end{IEEEkeywords}
%
\IEEEpeerreviewmaketitle

\textcolor{blue}{
}

\section{Introduction} \label{sec.intro}
Recently, energy demand has been increasing rapidly around the globe. According to reports by the Energy Information Administration, the world's energy demand is expected to witness a 50\% increase by 2050 compared to 2020 \cite{nalley2021ieo}. This increase must be met with a change in the current energy generation scheme which is done in a centralized fashion. Centralized energy generation facilities are located very far away from the end user, resulting in long transmission lines that lead to wasted energy. In addition, these facilities rely basically on fossil fuel power plants that have a huge environmental impact, such as air and water pollution and waste generation \cite{EPA_2024}.

To overcome these issues, a shift to a distributed \cite{hayyolalam2017qos} and renewable energy generation scheme is a viable solution. Distributed energy generation refers to various technologies that enable generating at or very close to the end user, which eliminates the wasted energy in the delivery process. In addition, it utilizes existing cost-efficient techniques to generate energy in residential areas using renewable energy sources. Thus, it increases the reliance on renewable energy sources, which leads to decreasing environmental hazards \cite{EPA_2023}.

\begin{figure}
    \centering
    \includegraphics[width=1\textwidth]{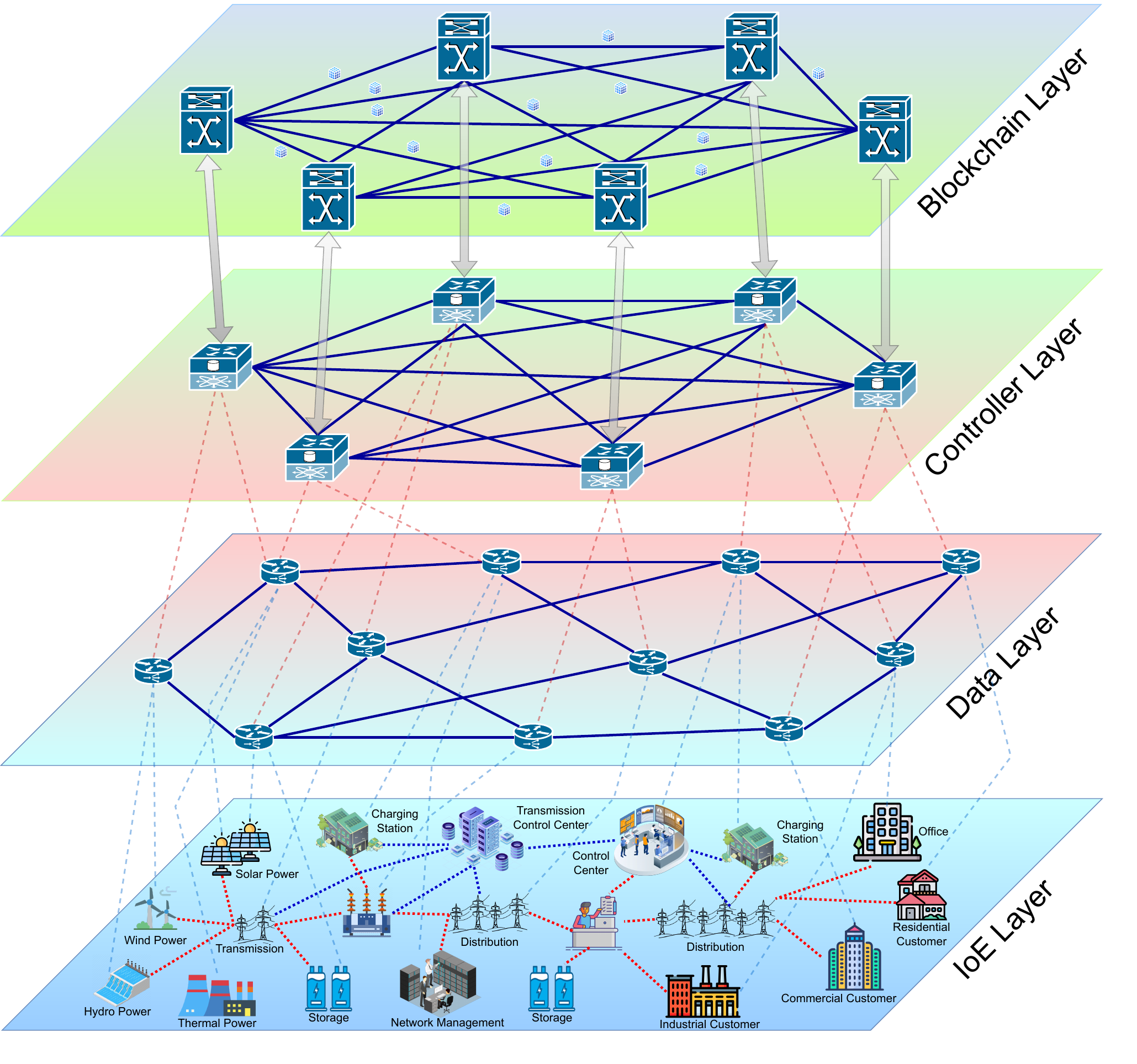}
    \caption{SDN and Blockchain-based architecture for the Internet of Energy.} \vspace{-.4cm}
    \label{fig.ioe_sdn_bc}
\end{figure}

However, to have a distributed system, one must address its challenges, including but not limited to security, reliability, and network performance. 
Creating a decentralized data-sharing architecture utilizing Software-Defined Network (SDN)-based infrastructure with Blockchain (BC) technology can be beneficial when there are numerous energy-producing entities, all managed through the Internet to reduce both energy consumption and wasted energy in an Internet of Energy (IoE) framework. In fact, the IoE is a technology that upgrades and automates the electric facilities for energy producers. This enables energy production to proceed more cleanly and efficiently with the least possible waste.
Figure \ref{fig.ioe_sdn_bc} illustrates such an architecture to enable efficient management of the IoE plane by decoupling control and data layers while ensuring secure and transparent data sharing, using the blockchain, among the stakeholders. \textcolor{black}{The figure depicts four key layers: (1) The IoE Layer, where different IoE devices exist, such as consumers, prosumers (consumers that sell energy), and distributed energy generators; (2) the Data Layer, which represents the aggregated data obtained from the IoE layer and the commands sent by the Control Layer to the IoE Layer; (3) the Control Layer, which manages the IoE Layer based on the data from the Data Layer, aiming to optimize energy flow and balance supply and demand; and (4) the Blockchain Layer, which ensures the security and transparency of energy transactions and, in some cases, enhances the security of the controllers in the control layer.} 

The main contributions of this work are as follows:
    
\begin{itemize}
    \item Analysis of the current state-of-the-art that integrates blockchain and SDNs into IoE. 
    \item Classification of research thrusts that use blockchain and SDNs in IoE based on the integration type, blockchain usage, and SDN usage.
    \item Articulation of different benefits of blockchain and SDNs into IoE.
    \item Identification of promising future work in the usage of blockchain and SDN in IoE. 
\end{itemize}

The rest of the paper is organized as follows. Enabling technologies related to BC-SDN-assisted solutions for IoE are introduced in Section \ref{sec.entech}.  Section \ref{sec.related} discusses the relevant state-of-the-art research works. Solutions that utilize both blockchain and SDN for the IoE systems are discussed in Section \ref{sec.sys}. The findings and discussion are provided in Section \ref{sec.experiments}. Section \ref{sec.future} points out the lessons learned from the literature and highlights the future trends in this topic. Finally, Section \ref{sec.conclusion} states the conclusion of this research. \\



\section{Enabling Technologies} \label{sec.entech}
This section addresses the essential enabling technologies in the field of BC-SDN-assisted solutions for IoE. 


\subsection{Software-Defined Networking}
Traditional network configuration is hardware-based and handled in a distributed manner, where each router should be configured separately. The combination of data and control planes has led to complex and hard-to-configure routers. More than that, incompatibility between different routers has led to an additional burden on the network administrators. Such difficulties led to the Software-Defined Networking (SDN). In SDNs, the separation between the data plane and control plane has led to simpler routers to be used at the data plane while having a centralized controller to manage and monitor all the network remotely \cite{haji2021comparison}. Thus, the previously mentioned problems of traditional networking are solved by SDNs. 

\subsection{Blockchain and Smart Contracts}
Blockchain technology leverages cryptography to maintain a shared, continuously growing database of records called a ledger, preserving the network's state. Decentralization in the blockchain is achieved through consensus protocols, such as PoW (Proof-of-Work) and PoS (Proof-of-Stake), which are sets of rules followed by network participants to agree on what constitutes a valid entry in the ledger. Blockchain eliminates the need for centralized governance and authority, enhancing the system's reliability and reducing latency across the network \cite{zheng2017overview}. 

Smart contracts refer to self-verifiable programs that run on the blockchain in a decentralized and transparent manner \cite{mohanta2018overview}. A smart contract represents individuals' reliance on code in the absence of trust between them, as trust is placed within the contract itself.

\subsection{Smart Grids and Internet of Energy}
The current energy scheme is a centralized solution where energy consumers exchange energy with the centralized utility grid. However, the increase in energy needs led to the emergence of distributed energy resources. As a result, the utility grid is no longer the only source of energy, and energy can be traded between the consumers and those distributed resources, where the term \emph{smart grid} appears. In Smart Grids (SG), energy is traded between the stakeholders over the network \cite{bui2012internet}. IoE could be seen as the integration of SGs with the Internet of Things (IoT) technologies \cite{ghiasi2023evolution} where SGs, distributed energy resources, consumers, and electric vehicles trade energy, share information, and get monitored or controlled through the Internet \cite{shahinzadeh2019internet}. \\

\section{Related Work} \label{sec.related}
This section reviews the relevant existing survey works. Then, their key aspects are compared side-by-side. Eventually, the motivation for conducting this novel survey research is stated.

Researchers in \cite{turner2023promising} have surveyed the integration of SDN and BC for IoT Networks. They have addressed the overview of SDN and BC technologies, their individual advantages and disadvantages, and the benefits and challenges of integrating them into IoT-based networks. They have introduced a novel classification of relevant research works with respect to their aim of implementation and research gaps concerning trust management, infrastructure, data management, and security. A thorough discussion of current works, whose aim is proposing an IoT infrastructure by the integration of the development and deployment of SDN and BC, has been presented. Finally, the authors have stated the open research avenues and possible future roadmaps for the combination of SDN and BC for IoT applications. 

Moreover, the authors in \cite{bhuyan2022survey} have surveyed the integration of blockchain, SDN, and Virtual Network Function (NFV) for smart home security. To this end, they have assessed the existing works relevant to smart home security. Also, they have provided a thorough and structured outline of the attacks and vulnerabilities in smart homes, and security solutions for smart homes via BC, SDN, and NFV technologies. The security performance assessment for smart homes has been presented, and the authors have also highlighted the merits of adopting BC, SDN, and NFV for smart home security. In addition, the simulation tools, adopted datasets, and evaluation metrics have been discussed. 

Authors in \cite{varma2023comprehensive} have proposed a survey with a concentration on various SDN control-plane architectures, security services, and challenges in SDN-based Vehicular Networks (SDVNs). They have highlighted the existing blockchain-oriented methods concentrating on security concerns in vehicular networks. Besides their particular architectures, the authors have addressed the components of the Internet of Vehicles (IoV), vehicular communication technologies, SDN, and blockchain. They have proposed a taxonomy of SDN-oriented security services in Vehicular Ad-hoc NETworks (VANETs) and different attacks on SDVN, accompanying solutions specific to their domain. The simulation tools, SDN controllers, and blockchain platforms for VANETs have also been highlighted. 

The study in \cite{aggarwal2021survey} has provided an extensive review of the essential concepts related to energy trading and the implication of empowering technologies adopted for dealing with energy exchanges through smart grids. They have categorized the existing energy trading solutions in the smart grid into three categories, including incentive, simulation, and mathematical models. The researchers have also classified the existing solutions according to their enabling technologies, such as SDN, IoE, and BC. Eventually, the roadmaps for upcoming research works in this area have been highlighted to help future researchers propose robust energy trading strategies in the smart grid. Although this survey has addressed SDN, BC, and IoE technologies, it has not focused on the studies that include all these three technologies in their proposed strategies.

Furthermore, authors in \cite{guo2022blockchain} have reviewed the existing research works, which intend to integrate emerging BC technology with smart grids. Their survey has explored the need for implementing blockchain technology in various aspects of smart grids. It aims to address the uncovered challenges in the existing solutions and shed light on the methodologies and strategies employed to merge blockchain with smart grid infrastructure. They have also provided a comprehensive comparison of blockchain-based solutions for smart grids from different viewpoints to address intuitions on the integration of smart grids and blockchain for different smart grid management tasks. Eventually, they have listed the unaddressed issues and highlighted potential future roadmaps in this field.

Table \ref{tab.rel} summarizes and compares the state-of-the-art relevant surveys side-by-side, emphasizing their focus on the concepts of BC, SDN, and IoE. Moreover, their application domains (e.g., SG, P2P Energy trading, and Electric vehicles) are highlighted. According to the observations, the studies in the literature address the following: blockchain technology application on SDN \cite{turner2023promising, bhuyan2022survey, varma2023comprehensive}; blockchain for smart grid \cite{guo2022blockchain}; energy trading in smart grids \cite{aggarwal2021survey, guo2022blockchain}; blockchain-based energy trading \cite{guo2022blockchain}. 
To the best of our knowledge, there is no existing research that thoroughly addresses the integration of Blockchain and SDN within the Internet of Energy domains (energy trading, smart grids, and electric vehicles). \\ 

\begin{table*}[htpb!]
    \centering
    \caption{Comparison of existing surveys.}
    ($^*$Although SDN, BC, and IoE are addressed, the focus is not on the studies that include all these technologies in their proposed strategies.)
   
    \begin{tabular}{c c c c c c c c}\\ 
     \hline
         \textbf{Ref} & \textbf{BC}& \textbf{SDN} & \textbf{Smart grid} & \textbf{P2P Energy trading} & \textbf{Electric vehicles} & \textbf{Application domain}\\
    \hline
        \cite{turner2023promising} & \Checkmark & \Checkmark & \ding{53} & \ding{53} & \ding{53} &  IoT\\
        \cite{bhuyan2022survey} & \Checkmark & \Checkmark & \ding{53} & \ding{53}  & \ding{53}  & Smart home security\\
        \cite{varma2023comprehensive} & \Checkmark & \Checkmark & \ding{53} & \ding{53}  & \Checkmark & Vehicular networks\\
        \cite{aggarwal2021survey} & \Checkmark $^*$ & \Checkmark $^*$ & \Checkmark & \Checkmark  & \ding{53}  & Smart grids\\
        \cite{guo2022blockchain} & \Checkmark & \ding{53} & \Checkmark & \Checkmark   & \ding{53} & Smart grids\\
        Ours & \Checkmark & \Checkmark & \Checkmark & \Checkmark & \Checkmark & 
        IoE (ET, SG, EV) 
        \\
    \hline
         
    \end{tabular}
    
    \label{tab.rel}
\end{table*}  

   
         
    

\section{Synergies of Blockchain and SDN in IoE}  \label{sec.sys}
\begin{figure}[h]
    \centering
    \includegraphics[width=1.1\textwidth]{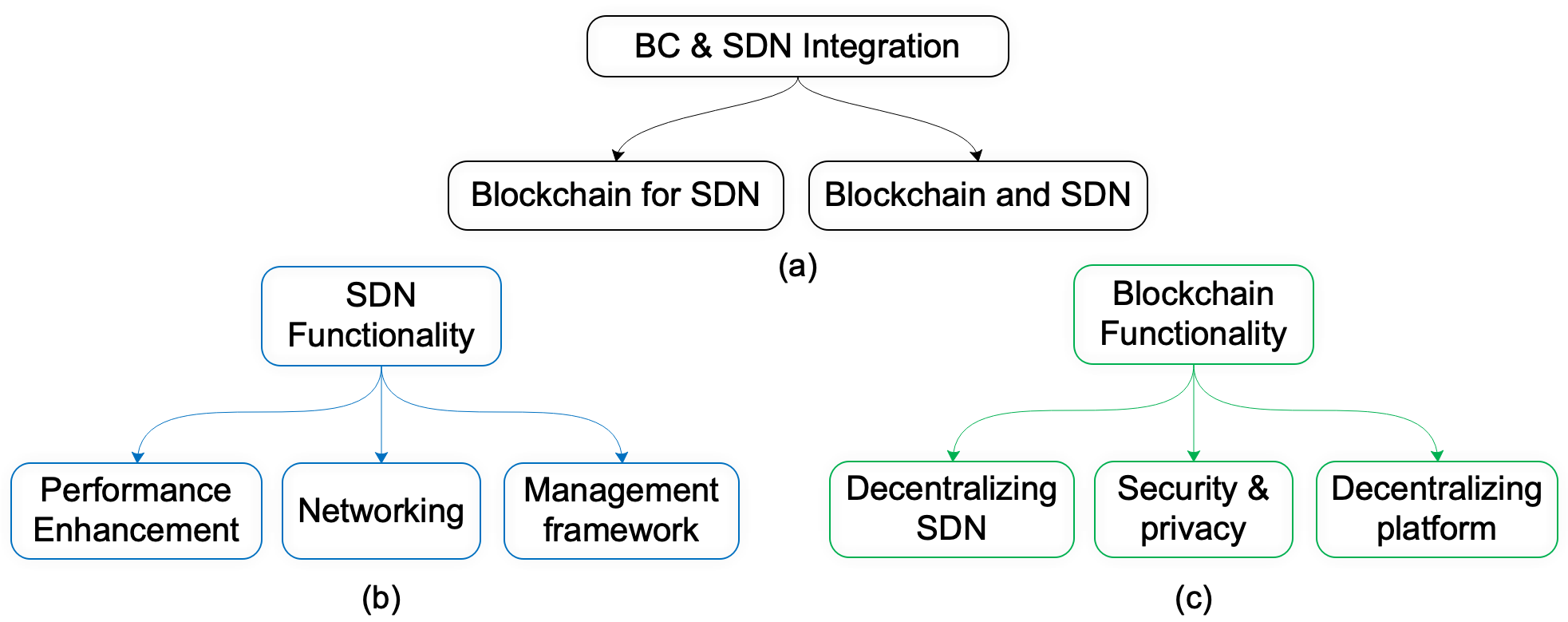}
    \caption{Classification Criteria for BC and SDN in IoE.}
    \label{fig:classificaiton1}
\end{figure}

In this section, we provide a summary of solutions that utilized both Blockchain and SDN in the IoE. Then, in Section \ref{sec.experiments}, we classify them based on three main aspects: a) Blockchain and SDN integration, b) SDN functionality, and c) Blockchain functionality, as shown in Figure \ref{fig:classificaiton1}.


In \cite{xiong2021distributed}, the researchers presented a solution called ClusterBlock that used a distributed SDN controller cluster to prevent a single point of failure and balance equipment load. Each SDN cluster had a cluster head functioning as a blockchain node, forming an SDN cluster head blockchain that enhances the security and privacy of the network. ClusterBlock also introduced a blockchain-based distributed control strategy and an algorithm for network attack detection, where the Jaccard similarity coefficient was utilized for attack detection. Simulation experiments comparing the ClusterBlock with existing OpenFlow protocol-based models showed that ClusterBlock offers more stable bandwidth and stronger security against DDoS attacks of similar magnitude. 
In summary, the usage of blockchain technology led to enhancing the privacy and security of the SDN communication network.
The proposed work could be considered as decentralizing the management of SDNs using blockchain technology and a consensus protocol to defend against possible attacks while getting rid of the single point of failure problem. To reduce the rate of communication with the blockchain, only the head of the cluster will be dealing with it, where the blockchain holds the latest flow table. After a network node sends a request to a cluster, upon the success of the request, the cluster header will store the IDs of the SDN domain and the node on-chain.
Their proposed solution was tested with a traditional smart grid setup involving automated substation control and communication infrastructure and network layout aligned with the IEC 61850 standard.

Researchers in \cite{bhattacharya2022evblocks} have designed a trusted and secure Energy Trading (ET) strategy for EVs with a focus on a vehicle-to-anything (V2X) environment. The proposed strategy, called EVBlocks, smooths the path for ET among different entities (e.g., smart grids, charging stations, and electric vehicles) in a trusted and secure way via a permissioned blockchain. EVBlocks is conducted in three phases. Firstly, the researchers have addressed the ET service deployed over an SDN-based 5G network with the aim of enabling a durable and real-time network arrangement for V2X nodes. The SDN-based 5G V2X network enables V2X nodes to eliminate third parties and cope with various demands in a short response time. The second phase aims to optimize the cost function and facilitate convergence to obtain at least one Nash equilibrium point by utilizing a non-cooperative game. During the final phase, the authors proposed a consensus mechanism called Proof-of-Greed (PoG) to deal with the changes in charging or discharging the EVs via an event-based scheduling method. The experimental results have been compared to the existing research works and proved the superiority of the proposed strategy.


Authors in \cite{kumar2023digital} have incorporated Digital Twin (DT), Deep learning (DL), BC, and SDN into the SG network to address the vulnerability of SG systems to different security threats since those traditional systems mostly trade data and services through insecure channels that are public. The researchers first developed a secure channel for communication leveraging BC, which provides an authentication strategy with the capability to resist in front of various famous assaults. Then, they developed a new DL network with fully connected layers, employing softmax to improve the SG network performance by a self-attention technique for attack detection. Afterward, they leveraged the SDN technology as the backbone of the system’s network to diminish the latency of the system and enhance the quality of service and the possibility of real-time services. Eventually, they combined the SDN control plane with the DT, which keeps the states of operations and behavior models of communications. The experimental results demonstrate that the proposed method provides an efficient authentication mechanism and a secure intrusion detection technique.

Furthermore, researchers in \cite{de2021peer} introduced a blockchain-based energy trading framework to be operated in an islanded part of the grid. The proposed solution consists of two main modules: auction and anomaly detection, both of which are enabled in a decentralized manner through smart contracts. Though blockchain provides transparency, traceability, and immutability, it does not provide a way of verifying data correctness; thus, anomaly detection was implemented. Another reason for using the blockchain is to enhance the privacy of the system by making the e-auction module based on the smart contract. The usage of a private blockchain led to an increase in the privacy of the proposed solution. An ERC20 token was used as a payment method in the proposed system.

The SDN technology enhances ubiquitous networks with programmable controllers and enables separation of the control plane and data plane, which increases flexibility. Such a control paradigm facilitates the segregation of distributed energy devices from their applications, forming a Software Defined Energy Internet (SDEI). To address the problems of credit crises and not being able to attract contributors in the energy trading markets, researchers in \cite{cao2022hierarchical} proposed an SDN and blockchain-based solution. Blockchain technology records the transactions and provides trusted services without a central authority. Specifically, blockchain manages and stores energy trading information among interested entities, while smart contracts within the blockchain reduce credit costs effectively for the stakeholders in the market.
The usage of SDN controllers can be seen in the devices used by operators in their edge control layer. 
The controllers collaborate to manage and direct the energy-conveying process within the physical energy layers. Each server is a node within the permissioned consortium blockchain, performing tasks such as bookkeeping, broadcasting, consensus, and verification. The consortium BC offers reliable and consistent intermediary services for transacting energy in the industrial application layer.

In \cite{lu2019blockchain}, the researchers proposed an SDN-based distributed energy trading scheme that utilized blockchain technology and focused on privacy protection. 
By applying SDN concepts and methods to the IoE, operations, control, and management can be partially or fully separated, leading to a more flexible IoE structure and operations. However, this SDN-based IoE revolution also presents challenges, notably in handling the substantial data volume and ensuring security and privacy in distributed systems.
While some blockchain-based schemes for energy trading exist, they prioritize data integrity without addressing trading matching while ensuring data security and privacy. The proposed scheme integrates blockchain technology with the SDN-based IoE, offering a distributed energy trading solution that ensures secure, reliable, and efficient electrical energy trading in a distributed environment.

Moreover, in \cite{chaudhary2019best}, a secure Electric Vehicles (EVs)  energy trading system is proposed, leveraging blockchain technology to ensure resilience against potential single points of failure. The blockchain is utilized to validate EVs' requests in a decentralized manner, with miner nodes chosen based on criteria such as energy requirements, duration of stay, dynamic pricing, connectivity record, and operational needs. Employing an SDN framework as the network backbone, leading to facilitating the transfer of EVs' requests to a global SDN controller, assisted in achieving low latency and real-time services. The adoption of the consortium blockchain within the proposed system is crucial for managing secure identities and transactions, recording events, and tracing faults.


The study in \cite{grammatikis2021sdn} introduced an SDN-based architecture aimed at enhancing the resilience of smart grids while also addressing intrusion detection and mitigation. The proposed solution leverages SDN technology across three key modules: the SDN-microSENSE Risk Assessment Framework (S-RAF), the Cross-Layer Energy Prevention and Detection System (XL-EPDS), and the SDN-enabled Self-healing Framework (SDN-SELF). These modules are strategically employed across the SDN planes, where they interact with SDN controllers to detect and mitigate intrusions. The SDN-SELF module plays a dual role, first by handling anomalies and intrusions identified by the XL-EPDS module and secondly by managing and optimizing energy usage. Notably, the energy management aspect incorporates a blockchain-based energy trading system for securing transactions, consisting of two key modules: an e-auction module and an intrusion and anomaly detection module.

Besides, in \cite{jindal2019survivor}, researchers presented a system model for energy trading between EVs and Charging Stations (CSs). It leveraged an edge-as-a-service framework to handle EV requests closer to their physical location in order to reduce latency and increase throughput. Moreover, two different utility functions are used to ensure efficient energy trading in the system. The first utility function is used for edge node selection, taking into account the delay, throughput, and the physical distance the EV has to travel to recharge. Then, once an edge node is selected, the other utility function is considered, which is related to the energy trading process. Considering EVs and CSs as buyers and sellers, the objective of the utility function is to maximize EVs' energy levels and CSs' profits, taking into account the energy levels available and the prices. Furthermore, a public blockchain using PoW consensus protocol is proposed to ensure the security and integrity of the network and the transactions. For a single transaction, the edge nodes that are not responsible for initiating the transaction act as the approver nodes, and once more than 50\% of them approve the transaction, then it is added to the blockchain. 


    

\section{Findings and Discussion} \label{sec.experiments}
By investigating the usage of both blockchain and SDNs in the Internet of Energy field, we identified the studies that fall into three Internet of Energy subdomains: Energy Trading \cite{cao2022hierarchical, lu2019blockchain, de2021peer}, Smart Grids \cite{xiong2021distributed, grammatikis2021sdn, kumar2023digital}, and Electric Vehicles \cite{bhattacharya2022evblocks, chaudhary2019best, jindal2019survivor}, as summarized in Table \ref{tab:summary}.

\subsection{Classification based on BC and SDN integration}
The studied works that both include blockchain and SDN could be classified into two categories: (1) Blockchain for SDN, and (2) Blockchain and SDN.

\subsubsection{Blockchain for SDN}\label{bc4sdn}  \vspace{-.2cm}
In this category, the blockchain is used to provide a service to the SDN itself or to enhance it, as in \cite{bhattacharya2022evblocks, xiong2021distributed} where the blockchain was used to decentralize the control plane in the SDN. In \cite{xiong2021distributed}, the blockchain prevented the single point of failure problem by utilizing blockchain-based control clusters. Besides, in \cite{bhattacharya2022evblocks}, BC is utilized to deal with the fluctuations in charging/discharging EVs via an event-based scheduling method. 

\subsubsection{Blockchain and SDN}  \vspace{-.2cm}
In this category, the blockchain and SDN were used together in the proposed solution without having one of them provide a service to the other directly. In other words, each technology provides an added value to the proposed system, and the blockchain does not provide any benefit to SDN directly, nor does the SDN to blockchain directly. The works classified under this category are \cite{de2021peer, cao2022hierarchical, lu2019blockchain, chaudhary2019best, grammatikis2021sdn, jindal2019survivor, kumar2023digital}.

\subsection{Classification based on SDN functionality}
Based on the SDN functionality, we found that it was used for three different purposes.

\subsubsection{Performance Enhancement}
In \cite{bhattacharya2022evblocks,cao2022hierarchical, jindal2019survivor, grammatikis2021sdn, chaudhary2019best}, SDNs were used to enhance performance. More specifically, SDN was used to minimize response time by enabling V2X without intermediaries, as we can see in \cite{bhattacharya2022evblocks}. In \cite{chaudhary2019best}, it was mentioned that the SDN plays a role in reducing network latency and enhancing the Quality of Service (QoS). Researchers in \cite{jindal2019survivor} indicated that the usage of SDN for provision of dynamic network policies will reduce the latency, thus the performance of the entire system is increased. In \cite{grammatikis2021sdn}, the increase in smart grid resilience was noted as a benefit of using SDNs. Finally, the reduction in network latency was a benefit of using SDN in \cite{kumar2023digital}.

\subsubsection{Networking}  \vspace{-.2cm}
The SDN was used instead of the traditional network, as noticed in \cite{xiong2021distributed, de2021peer} where the researchers did not specify any special reasons for using SDN in their proposed solutions.
 
\subsubsection{Management Framework}  \vspace{-.2cm}
In \cite{cao2022hierarchical, lu2019blockchain}, it was noticed that the logic of SDN in separating the control plane from the data plane was adapted as a management framework. In \cite{lu2019blockchain}, the SDN's concept and methods are applied to the IoE to separate the operation, control, and management of the IoE. In \cite{cao2022hierarchical}, the SDN concept was adapted to separate distributed energy devices from energy applications flexibly. 

\subsection{Classification based on BC functionality}  \vspace{-.2cm}
Three different purposes for utilizing blockchain were found as follows.

\subsubsection{Decentralizing SDN}  \vspace{-.2cm}
As noticed in \cite{xiong2021distributed, bhattacharya2022evblocks}, the blockchain enabled the decentralization of SDN controllers, where those researches are the same we classified in subsection \ref{bc4sdn}. 

\subsubsection{Security and Privacy}  \vspace{-.2cm} In \cite{jindal2019survivor, lu2019blockchain, xiong2021distributed}, blockchain was utilized to enhance security and privacy by securing the energy trading transactions, securing communications, and preserving privacy. In \cite{de2021peer}, the blockchain had a role in overseeing the security condition of smart grid devices. In \cite{chaudhary2019best}, the blockchain had a role in defending against man-in-the-middle-attacks (MiTM). In \cite{grammatikis2021sdn}, a smart contract-based module was introduced for intrusion and anomaly detection.

\subsubsection{Decentralizing Platform}  \vspace{-.2cm} In this usage, the blockchain's smart contract was used to enable energy trading and the matching between consumers and prosumers in a decentralized manner. In \cite{cao2022hierarchical}, the information related to energy trading between interested entities is stored and managed on the blockchain in a reliable manner, along with a trading smart contract. In \cite{grammatikis2021sdn}, an e-auction smart contract-based module was developed for energy trading, and another smart contract-based module was developed to enhance security.

The three types of blockchain (public, consortium, and private) were noted in the studied works. As the consensus protocol, practical Byzantine Fault Tolerance (pBFT), Kafka, Proof-of-Work (PoW), Proof-of-Greed (PoG), and a custom voting-based consensus were observed. Table \ref{tab:summary} summarizes the type of integration method, the blockchain usage, the SDN usage, the blockchain type, and the consensus protocol for each of the investigated studies. \textcolor{black}{ 
 The inspected papers are categorized into three groups based on their application domains, including energy trading, smart grids, and electric vehicles (Table \ref{tab:summary}, column "Domain"). From another viewpoint, considering the SDN functionality, the reviewed papers are classified into three main groups, namely management framework, networking, and performance enhancement (Table \ref{tab:summary}, column "SDN Usage"). \\ 
 }
\begin{table*}[htpb!]
    \centering
    \caption{Summary of the domain, BC usage, SDN usage, blockchain type, and consensus protocol of the solutions.}

    \begin{tabular}{|c|c|c|c|c|c|}
    \hline
         \textbf{Domain} & \textbf{Ref} & \textbf{BC Usage} & \textbf{SDN Usage} & \textbf{Blockchain Type} & \textbf{Consensus Protocol}\\
        \hline
         \multirow{3}{*}{Energy Trading} & \cite{cao2022hierarchical} & Security & \multirow{2}{*}{Management Framework} & Consortium & pBFT \\
         \cline{2-3} \cline{5-6}
         & \cite{lu2019blockchain} & Decentralizing Platform & & - & -\\
         \cline{2-6}
         & \cite{de2021peer} & Security &\multirow{2}{*}{Networking}& Private & Kafka\\
         \cline{1-3} \cline{5-6}
         \multirow{3}{*}{Smart Grids} &  \cite{xiong2021distributed}  & Decentralizing SDN & & Public & PoW \\
         \cline{2-6}
         & \cite{grammatikis2021sdn}  & Decentralizing Platform & \multirow{5}{*}{\parbox[c]{2cm} {{Performance Enhancement}}} & Permissioned & Kafka\\
         \cline{2-3} \cline{5-6}
         & \cite{kumar2023digital} & Security & & Consortium & PoG \\
         \cline{1-3} \cline{5-6}
         \multirow{3}{*}{Electric Vehicles}  & \cite{bhattacharya2022evblocks} & Decentralizing SDN & & Public & Voting-based Consensus \\
         \cline{2-3} \cline{5-6}
         & \cite{chaudhary2019best} & Security & & Consortium & PoW\\
         \cline{2-3} \cline{5-6}
         & \cite{jindal2019survivor} & Security & & Public & PoW \\
         \hline
    \end{tabular}
    \label{tab:summary}
\end{table*}

\section{Lessons Learned and Future Trends} \label{sec.future}

This section includes two parts. The first part describes the important lessons learned from the literature, and the second part discusses the open issues.

\subsection{Lessons Learned}  \vspace{-.2cm}
    This section highlights the essential challenges and lessons learned from a comprehensive analysis of the state-of-the-art works.
    
\subsubsection{Decentralization and P2P Trading}  \vspace{-.2cm}
    The integration of SDN and BC in IoE enables ET systems to facilitate P2P trading by decentralized control and decision-making without intermediaries. This can empower stakeholders, enhance the adaptability of the energy grid, and prevent cyber-attacks or single points of failure.

\subsubsection{Transparency Improvement}
    The integration of immutable ledgers and the transparency of BC with SDN technology allows superior energy transaction tracking and data flow in ET systems and enhances accountability and transparency.

\subsubsection{Scalability and Performance}
    According to the literature, both SDN and BC individually present scalability challenges, while their integration paves the way for diminishing this problem and enhancing the ET system's overall performance via optimizing transaction processes and resource utilization.

\subsubsection{Data Privacy and Identity Management}
    Privacy-preservation of the participant data and guaranteeing secure identity management are very important in ET systems. The SDN and BC technologies provide solutions, such as permissioned blockchain, encryption, and decentralized identity management systems, to highlight these issues.

\subsubsection{Efficient Resource Management}
    The synergy between SDN and BC contributes to more effective resource management in ET systems through energy transaction optimization, dynamic network traffic management, and real-time supply and demand balancing.

\subsubsection{Integration with Renewable Energy resources}
    ET systems equipped with the combination of SDN and BC technologies have the capability of facilitating the combination of renewable energy resources via empowering 
    efficacious P2P trading and incentivizing renewable energy consumption and generation. Existing solutions offer different mechanisms to incentivize renewable energy generation and guarantee grid stability in decentralized energy trading systems.
    
\subsection{Future Trends and Challenges}
\begin{figure}[htbp!]
    \centering 
    \includegraphics[width=1\textwidth]{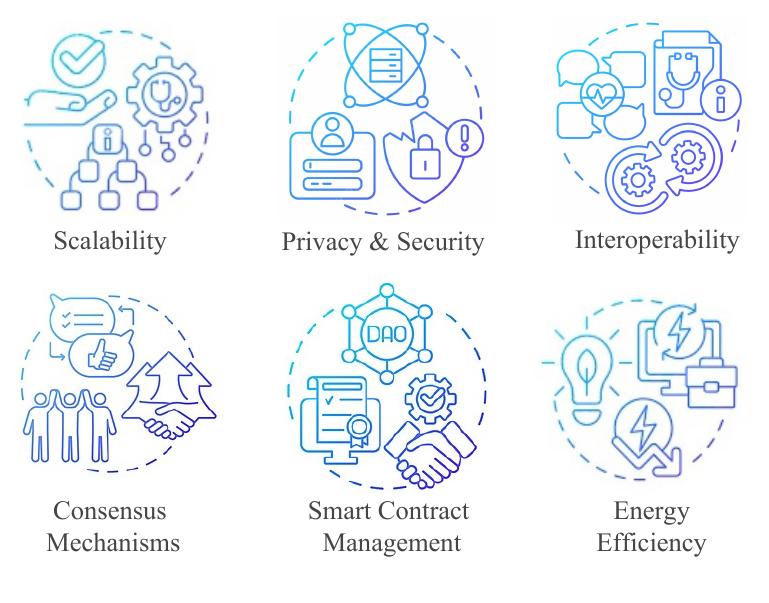} \vspace{-.8cm}
    \caption{A high-level overview of the future research directions to integrate BC and SDN for IoE.} \vspace{-.2cm}
    \label{fig.fut}
\end{figure}

    The integration of SDN and BC technology for ET systems offers promising opportunities while bringing out various challenges and open issues. Despite several research works that have been done in the field of BC-SDN-assisted IoE solutions, there are still uncovered aspects that need to be addressed. This section highlights and discusses some of these open challenges as a first step towards a roadmap to 
    draw more attention from the research community and trigger new and fresh ideas. 
    Figure \ref{fig.fut} illustrates the highlighted open issues graphically.

\subsubsection{Scalability}
    Since both BC and SDN need remarkable computational resources, scalability becomes a crucial problem in extensive ET systems. Thus, guaranteeing that the system has the capability of handling numerous transactions while preserving the performance and efficiency of the system is a vital challenge.

\subsubsection{Privacy and Security}
    ET systems usually involve sensitive information relevant to the patterns of consumption, user identities, and pricing. Guaranteeing transparency and public verifiability while preserving security and privacy is a vital challenging issue. In order to preserve the sensitive data as well as the integrity of the system, proposing robust cryptographic strategies and designing powerful techniques to maintain privacy is crucial.

\subsubsection{Interoperability}
    The integration of SDN and BC technologies usually involves various platforms, protocols, and standards. One of the critical challenges in this regard is to obtain interoperability across these assorted systems. Guaranteeing interoperability, seamless communication, and data exchange between diverse platforms and networks while maintaining the security and integrity of transactions is of utmost importance. 

\subsubsection{Consensus Mechanisms}
    Selecting a proper consensus mechanism for the BC-based network is so critical. In order to deploy real-time trading demands, ET systems need swift transaction processes. However, due to the inherited latency of conventional blockchain consensus mechanisms, they may not be appropriate for real-time energy trading systems. Developing efficient consensus mechanisms customized for ET systems is an open challenge in this domain.  
    
\subsubsection{Smart Contract Management} \vspace{-.2cm}
    Automating and imposing the terms of IoE agreements relies on smart contracts. However, guaranteeing the security, efficiency, and reliability of smart contracts is one of the critical challenges in this field. Sensitivity and susceptibility in smart contract code may contribute to financial losses and security breaches. Designating robust strategies for smart contract auditing, verification, and execution is demanded to reduce these risks.

\subsubsection{Energy Efficiency}
    Both SDN and BC technologies utilize remarkable amounts of energy. Energy consumption management is an important challenge, particularly regarding sustainable ET systems. Developing energy-efficient algorithms and mechanisms for diminishing the energy consumption of SDN controllers and BC consensus mechanisms is very important for minimizing environmental footprint and operational costs.


\section{Conclusion} \label{sec.conclusion}
Under the increasing demand for energy and the transformation of energy generation from highly centralized to more distributed sources with highly intermittent alternatives, unprecedented challenges are emerging. In this study, we have considered an emerging combination of BC and SDN to enhance the capabilities and opportunities of the Internet of Energy systems. 
By investigating the integration of SDNs and BC technologies within the domain of the Internet of Energy, 
we posit that such an integration enhances performance, improves security, enables decentralized operations, and eliminates single points of failure. The full potential of such synergistic integration seems to remain untapped. We categorize the solutions that leverage both SDNs and blockchain into two groups based on SDN and BC integration methods. Blockchain applications are noted to be decentralizing the SDN control plane, serving as a decentralized platform, and enhancing security, while SDNs contributed as performance enhancers, substitution for traditional networking, and a logical management framework to separate the control plane and energy routing plane. 


\ifCLASSOPTIONcompsoc
\else
\fi

\section*{Acknowledgment}
This work is supported by TÜBITAK (The Scientific and Technical Research Council of Türkiye) 2247-A National Outstanding Researchers Program Award 121C338. 




%



\end{document}